\documentstyle[12pt]{article}
\newcommand{\bq}{\begin{equation}}
\newcommand{\eq}{\end{equation}}
\newcommand{\ba}{\begin{eqnarray}}
\newcommand{\ea}{\end{eqnarray}}

\newcommand{\nl }{ \nonumber  }

\newcommand{\ul}{\underline}
\newcommand{\p}{\partial}
\newcommand{\pu}{\p_\tau}
\newcommand{\pJ}{\p_J}
\newcommand{\pj}{\p_j}
\newcommand{\h}{\hspace{1cm}}
\newcommand{\s}{\sigma}
\newcommand{\uz}{\underline z}
\newcommand{\us}{\underline\sigma}
\newcommand{\da}{\delta^p(\us_1 - \us_2)}
\newcommand{\la}{\lambda}
\newcommand{\La}{\Lambda}
\begin{document}
%\pagestyle{empty,textwidth,textheight}
%%%\begin{titlepage}
%\begin{flushright}
%{\bf hep-th/99}
%\end{flushright}
\vspace*{.5cm}
{\bf\begin{center}
  D=10 CHIRAL TENSIONLESS SUPER p-BRANES 
\footnote{Work supported in part by the National Science Foundation
of Bulgaria under contract $\phi-620/1996$}
\vspace*{.5cm}
\\
P. Bozhilov
\footnote {E-mail: bojilov@thsun1.jinr.ru; permanent address:
Dept.of Theoretical Physics,"Konstantin Preslavsky" Univ. of 
Shoumen, 9700 Shoumen, Bulgaria}  \\
\it Bogoliubov Laboratory of Theoretical Physics, \\ 
JINR, 141980 Dubna, Russia  
\vspace*{.5cm} 
\end{center}}

%====abstract====
We consider a model for tensionless (null) super $p$-branes
with $N$ chiral supersymmetries in ten dimensional flat space-time.
After establishing the symmetries of the action, we give 
the general solution of the classical equations of motion in a 
particular gauge. In the case of a null superstring ($p$=1) we find 
the general solution in an arbitrary gauge. Then, using a harmonic 
superspace approach, the initial algebra of first and second 
class constraints is converted into an algebra of 
Lorentz-covariant, BFV-irreducible, first class constraints only. 
The corresponding BRST charge is as for a first rank dynamical 
system.  
%====end of abstract==== 

%%%\end{titlepage} 
%%%\normalsize 
\vspace*{.5cm} 

%%%%%%%%%%%%%%%%%%%%%%%%%%%%%%%%%%%%%%%%%%%%%%%%%%%%%%%%%%%%%%%%%%%%%%%%%%%%
\section{\bf Introduction}
%%%%%%%%%%%%%%%%%%%%%%%%%%%%%%%%%%%%%%%%%%%%%%%%%%%%%%%%%%%%%%%%%%%%%%%%%%%%
\hspace{1cm}
The tensionless (null) $p$-branes correspond to usual $p$-branes with their
tension $T_p$ taken to be zero. This relationship between null 
$p$-branes and the tensionful ones may be regarded as a 
generalization of the massless-massive particles correspondence. On 
the other hand, the limit $T_p \to 0$ corresponds to the 
energetic scale $E>>M_{Plank}$. In other words, the null $p$-brane 
is the high energy limit of the tensionful one. There exist also an 
interpretation of the null and free $p$-branes as theories, 
corresponding to different vacuum states of a $p$-brane, 
interacting with a scalar field background \cite{BZ}. So, one can
consider the possibility of tension generation for null $p$-branes
(see \cite{BSTV} and references therein). Another viewpoint on the connection
between null and tensionful $p$-branes is that the null one may be
interpreted as a "free" theory opposed to the tensionful "interacting"
theory \cite{G}. All this explains the interest in considering null
$p$-branes and their supersymmetric extensions.

   Models for tensionless $p$-branes with manifest supersymmetry
are proposed in \cite{Zh}. In \cite{BZ} a twistor-like action
is suggested, for null super-$p$-branes with $N$-extended global
supersymmetry in four dimensional space-time. In the recent work \cite{S}, 
the quantum constraint algebras of the usual and 
conformal tensionless spinning $p$-branes are considered.

In a previous paper \cite{136}, we announced for a null super
$p$-brane model, and here we are going to formulate it, and to 
consider its classical properties. After establishing the symmetries of 
the action, we give the general solution of the classical equations of 
motion in a particular gauge. In the case of a null superstring, 
($p$=1), we find the general solution in an arbitrary gauge. Then, 
in the framework of a harmonic superspace approach, 
the initial algebra of first 
and second class constraints is converted into an algebra of 
Lorentz-covariant, BFV-irreducible, first class constraints only. 
The corresponding BRST charge is as for a first rank dynamical 
system.  

%%%%%%%%%%%%%%%%%%%%%%%%%%%%%%%%%%%%%%%%%%%%%%%%%%%%%%%%%%%%%%%%%%%%%%%%%%%%
\section{\bf Lagrangian formulation}
%%%%%%%%%%%%%%%%%%%%%%%%%%%%%%%%%%%%%%%%%%%%%%%%%%%%%%%%%%%%%%%%%%%%%%%%%%%%
\hspace{1cm}
We define our model for $D=10$ $N$-extended chiral tensionless 
super $p$-branes by the action:
\ba\label{a}
S=\int d^{p+1}\xi L \h,\h
L=V^JV^K\Pi_J^\mu\Pi_K^\nu\eta_{\mu\nu},
\\ \nl
\Pi_J^\mu=\pJ x^\mu+i\sum_{A=1}^N (\theta^A\s^\mu\pJ\theta^A)\h,\h
\pJ=\p/\p\xi^J,
\\ \nl
\xi^J=(\xi^0,\xi^j)=(\tau,\s^j),\h diag(\eta_{\mu\nu})=(-,+,...,+),
\\ \nl
J,K=0,1,...,p \h,\h j,k=1,...,p \h,\h \mu,\nu=0,1,...,9.
\ea
Here $(x^\nu,\theta^{A\alpha})$ are the superspace coordinates,
$\theta^{A\alpha}$ are $N$ left Majorana-Weyl space-time spinors
($\alpha=1,...,16$ , $N$ being the number of the supersymmetries)
and $\s^\mu$ are the 10-dimensional Pauli matrices (our spinor 
conventions are given in the Appendix). Actions of this type are 
first given in \cite{Lind} for the case of tensionless superstring
($p=1,N=1$) and in \cite{CQG} for the bosonic case ($N=0$).

The action (\ref{a}) has an obvious global Poincar$\acute{e}$ 
invariance. Under global infinitesimal supersymmetry 
transformations, the fields $\theta^{A\alpha}(\xi)$, $x^\nu(\xi)$ 
and $V^J(\xi)$ transform as follows:
\ba\nl
\delta_\eta^{A\alpha}=\eta^{A\alpha}\h,\h
\delta_\eta x^\mu=i\sum_A(\theta^A\s^\mu\delta_\eta\theta^A)\h,\h
\delta_\eta V^J=0.
\ea
As a consequence $\delta_\eta\Pi_J^\mu=0$ and hence $\delta_\eta L
=\delta_\eta S=0$ also.

To prove the invariance of the action under infinitesimal 
diffeomorphisms, we first write down the corresponding 
transformation law for the (r,s)-type tensor density of weight $a$
\ba\nl
\delta_{\varepsilon}T^{J_1...J_r}_{K_1...K_s}[a]&=&
L_{\varepsilon}T^{J_1...J_r}_{K_1...K_s}[a]=
\varepsilon^L\p_L T^{J_1...J_r}_{K_1...K_s}[a]\\
\label{diff}
&+&
T^{J_1...J_r}_{KK_2...K_s}[a]\p_{K_1}\varepsilon^K+...+
T^{J_1...J_r}_{K_1...K_{s-1}K}[a]\p_{K_s}\varepsilon^K \\ \nl
&-&
T^{JJ_2...J_r}_{K_1...K_s}[a]\p_J\varepsilon^{J_1}-...-
T^{J_1...J_{r-1}J}_{K_1...K_s}[a]\p_J\varepsilon^{J_r} \\ \nl
&+&
aT^{J_1...J_r}_{K_1...K_s}[a]\p_L\varepsilon^L ,
\ea
where $L_\varepsilon$ is the Lie derivative along the vector field 
$\varepsilon$. Using (\ref{diff}), one verifies that if 
$x^\mu(\xi)$, $\theta^{A\alpha}(\xi)$ are world-volume scalars 
($a=0$) and $V^J(\xi)$ is a world-volume (1,0)-type tensor density 
of weight $a=1/2$, then $\Pi_J^\nu$ is a (0,1)-type tensor, 
$\Pi_J^\nu \Pi_{K\nu}$ is a (0,2)-type tensor and $L$ is a scalar 
density of weight $a=1$. Therefore, 
\ba\nl 
\delta_{\varepsilon}S=\int d^{p+1}\xi\p_J\bigl ( \varepsilon^J L\bigr ) 
\ea 
and this variation vanishes under suitable boundary conditions.

Let us now check the kappa-invariance of the action. 
We define the $\kappa$-variations of $\theta^{A\alpha}(\xi)$, 
$x^\nu(\xi)$ and $V^J(\xi)$ as follows:
\ba\label{k}
\delta_\kappa\theta^{A\alpha}=i\bigl(\Gamma\kappa^A\bigr)^\alpha=
iV^J\bigl(\not{\Pi_J}\kappa^A\bigr)^\alpha,\h
\delta_\kappa x^\nu=-i\sum_A(\theta^A\s^\nu\delta_\kappa\theta^A),
\\ \nl
\delta_\kappa V^K=2V^K V^L \sum_A(\p_L\theta^A\kappa^A).
\ea
Therefore, $\kappa^{A\alpha}(\xi)$ are left Majorana-Weyl 
space-time spinors and world-volume scalar densities of weight 
$a=-1/2$. 

From (\ref{k}) we obtain:
\ba\nl
\delta_\kappa\bigl(\Pi_J^\nu \Pi_{K\nu}\bigr)=-2i\sum_A\bigl[
\p_J\theta^A\not{\Pi_K}+\p_K\theta^A\not{\Pi_J}\bigr]
\delta_\kappa\theta^A
\ea
and
\ba\nl
\delta_\kappa L=2V^J \Pi_J^\nu \Pi_{K\nu}\bigl[\delta_\kappa V^K-
2V^K V^L\sum_A(\p_L\theta^A\kappa^A)\bigr] = 0 .
\ea

The algebra of kappa-transformations closes only on the equations 
of motion, which can be written in the form:
\ba\label{eqm}
\p_J\bigl(V^JV^K\Pi_{K\nu}\bigr)=0,\h
V^JV^K\bigl(\p_J\theta^A\not{\Pi_K}\bigr)_\alpha=0,\h
V^J \Pi_J^\nu \Pi_{K\nu}=0 .
\ea
As usual, an additional local bosonic world-volume symmetry is 
needed for its closure. In our case, the Lagrangian, and therefore 
the action, are invariant under the following transformations of 
the fields:
\ba\nl
\delta_\lambda\theta^A(\xi)=\lambda V^J\p_J\theta^A,\h
\delta_\lambda x^\nu(\xi)=-i\sum_A(\theta^A\s^\nu\delta_\lambda
\theta^A),\h
\delta_\lambda V^J(\xi)=0 .
\ea
Now, checking the commutator of two kappa-transformations, we find:
\ba\nl
[\delta_{\kappa_1},\delta_{\kappa_2}]\theta^{A\alpha}(\xi)&=&
\delta_\kappa\theta^{A\alpha}(\xi)+
\mbox{terms $\propto$ eqs. of motion} , \\
\nl
[\delta_{\kappa_1},\delta_{\kappa_2}]x^\nu(\xi)&=&
(\delta_\kappa+\delta_\varepsilon+\delta_\lambda)x^\nu(\xi)+
\mbox{terms $\propto$ eqs. of motion} , \\
\nl
[\delta_{\kappa_1},\delta_{\kappa_2}]V^J(\xi)&=&
\delta_\varepsilon V^J(\xi)+
\mbox{terms $\propto$ eqs. of motion} .
\ea
Here $\kappa^{A\alpha}(\xi)$, $\lambda(\xi)$ and $\varepsilon(\xi)$ 
are given by the expressions:  
\ba\nl 
\kappa^{A\alpha}=-2V^K\sum_B[(\p_K\theta^B\kappa_1^B)\kappa_2^{A\alpha}-
(\p_K\theta^B\kappa_2^B)\kappa_1^{A\alpha}],\\
\nl
\lambda=4iV^K\sum_A(\kappa_1^A\not{\Pi_K}\kappa_2^A)\h,\h
\varepsilon^J=-V^J\lambda .
\ea

We note that $\Gamma_{\alpha\beta}=\bigl(V^J\not{\Pi_J}\bigr)_
{\alpha\beta}$ in (\ref{k}) has the following property on the 
equations of motion 
\ba\nl 
\Gamma^2 = 0 .  
\ea 
This means, that the local kappa-invariance of the action indeed 
eliminates half of the components of $\theta^A$ as is needed.

For transition to Hamiltonian picture, it is convenient to rewrite 
the Lagrangian density (\ref{a}) in the form 
($\pu=\p/\p\tau, \p_j=\p/\p\s^j$):
\ba\label{L}
L=\frac{1}{4\mu^0}\Bigl [\bigl (\pu-\mu^j\pj\bigr )x+
i\sum_A\theta^A\s\bigl (\pu-\mu^j\pj\bigr )\theta^A\Bigr ]^2 ,
\ea
where
\ba\nl
V^J=\bigl(V^0,V^j\bigr)=\Biggl(-\frac{1}{2\sqrt{\mu^0}},
\frac{\mu^j}{2\sqrt{\mu^0}}\Biggr)
\ea
The equations of motion for the Lagrange multipliers $\mu^{0}$ and
$\mu^{j}$ which follow from (\ref{L}) give the constraints ($p_\nu$ 
and $p^A_{\theta\alpha}$ are the momenta conjugated to $x^\nu$
and $\theta^{A\alpha}$):  
\ba\label{T0,Tj} 
T_0=p^2\h,\h T_j=p_\nu\p_j x^\nu+\sum_A 
p^A_{\theta\alpha}\p_j\theta^{A\alpha}.  
\ea 
The remaining constraints follow from the definition of the 
momenta $p^A_{\theta\alpha}$:
\ba\label{D}
D_\alpha^A=-ip_{\theta\alpha}^A-(\not{p}\theta^A)_\alpha.
\ea

%%%%%%%%%%%%%%%%%%%%%%%%%%%%%%%%%%%%%%%%%%%%%%%%%%%%%%%%%%%%%%%%%%%%%%%%%%%%%
\section{\bf Hamiltonian formulation}
%%%%%%%%%%%%%%%%%%%%%%%%%%%%%%%%%%%%%%%%%%%%%%%%%%%%%%%%%%%%%%%%%%%%%%%%%%%%%
\hspace{1cm}
The Hamiltonian which corresponds to the Lagrangian density 
(\ref{L}) is a linear combination of the constraint (\ref{T0,Tj}) 
and (\ref{D}):
\ba\label{H0}
H_0=\int d^p\sigma\bigl [\mu^0 T_0+\mu^j T_j+
\sum_A\mu^{A\alpha} D^A_{\alpha}\bigr ]
\ea
It is a generalization of the Hamiltonians for the bosonic null 
$p$-brane and for the $N$-extended Green-Schwarz superparticle.

The equations of motion which follow from the Hamiltonian 
(\ref{H0}) are:
\ba\nl
(\pu-\mu^j\pj)x^\nu&=&2\mu^0p^\nu-\sum_A(\mu^A\s^\nu\theta^A),\h
(\pu-\mu^j\pj)p_\nu=(\pj\mu^j)p_\nu,\\
\label{em}
(\pu-\mu^j\pj)\theta^{A\alpha}&=&i\mu^{A\alpha},\h
(\pu-\mu^j\pj)p^A_{\theta\alpha}=(\pj\mu^j)p^A_{\theta\alpha}
+(\mu^A\not{p})_\alpha .
\ea
In (\ref{em}), one can consider $\mu^0$, $\mu^j$ and 
$\mu^{A\alpha}$ as depending only on $\us=(\s^1,...,\s^p)$, but 
not on $\tau$ as a consequence from their equations of motion.

In the gauge when $\mu^0$, $\mu^j$ and $\mu^{A\alpha}$ are 
constants, the general solution of (\ref{em}) is 
\ba \nl 
x^\nu(\tau,\us)&=&x^\nu(\uz)+\tau\bigl [2\mu^0p^\nu(\uz)-
\sum_A(\mu^A\s^\nu\theta^A(\tau,\us))\bigr ], \\
\nl
&=&x^\nu(\uz)+\tau\bigl [2\mu^0p^\nu(\uz)-
\sum_A(\mu^A\s^\nu\theta^A(\uz))\bigr ] \\
\label{gsp}
p_\nu(\tau,\us)&=&p_\nu(\uz) ,\h
\theta^{A\alpha}(\tau,\us)=
\theta^{A\alpha}(\uz)+i\tau\mu^{A\alpha} ,\\
\nl
p^A_{\theta\alpha}(\tau,\us)&=&p^A_{\theta\alpha}(\uz)+
\tau (\mu^A\s^\nu)_{\alpha} p_{\nu}(\uz) ,
\ea
where $x^\nu(\uz)$, $p_\nu(\uz)$, $\theta^{A\alpha}(\uz)$ and 
$p^A_{\theta\alpha}(\uz)$ are arbitrary functions of their arguments
\ba\nl
z^j = \mu^j\tau+\s^j .
\ea
In the case of null strings ($p=1$), one can write 
explicitly the general solution of the equations of motion in an
arbitrary gauge:  $\mu^0=\mu^0(\s)$, $\mu^1\equiv\mu=\mu(\s)$, 
$\mu^{A\alpha}=\mu^{A\alpha}(\s)$. This solution is given by
\ba
\nl
x^\nu(\tau,\s)&=&g^\nu(w)-2\int\limits_{}^\s\frac{\mu^0(s)}{\mu^2(s)}ds
f^\nu(w)+\sum_A\int\limits^\s\frac{\mu^{A\alpha}(s)}{\mu(s)}ds
\bigl [\s^\nu\zeta^A(w)\bigr ]_\alpha \\
\nl
&-&i\sum_A\int\limits^\s 
ds_1\frac{(\mu^A\s^\nu)_\alpha(s_1)}{\mu(s_1)} 
\int\limits^{s_1}\frac{\mu^{A\alpha}(s)}{\mu(s)}ds ,\\
\label{gs1}
p_\nu(\tau,\s)&=&\mu^{-1}(\s)f_\nu(w) ,\\
\nl
\theta^{A\alpha}(\tau,\s)&=&\zeta^{A\alpha}(w)-i\int\limits^\s
\frac{\mu^{A\alpha}(s)}{\mu(s)}ds ,\\
\nl
p^A_{\theta\alpha}(\tau,\s)&=&\mu^{-1}(\s)\Biggl [h^A_\alpha(w)-
\int\limits^\s\frac{(\mu^A\s^\nu)_\alpha(s)}{\mu(s)}ds f_\nu(w)
\Biggr ] .
\ea
Here $g^\nu(w)$, $f_\nu(w)$, $\zeta^{A\alpha}(w)$ and 
$h^A_\alpha(w)$ are arbitrary functions of the variable 
\ba\nl 
w = \tau + \int\limits^\s \frac{ds}{\mu(s)} 
\ea
The solution (\ref{gsp}) at $p=1$ differs from (\ref{gs1}) by 
the choice of the particular solutions of the inhomogenious 
equations.  As for $z$ and $w$, one can write for example ($\mu^0$, 
$\mu$, $\mu^{A\alpha}$ are now constants) 
\ba\nl 
p_\nu(\tau,\s)=\mu^{-1}f_\nu(\tau+\s/\mu)=\mu^{-1}f_\nu[\mu^{-1}
(\mu\tau+\s)]=p_\nu(z)
\ea
and analogously for the other arbitrary functions in the general 
solution of the equations of motion.

Let us now consider the properties of the constraints 
(\ref{T0,Tj}), (\ref{D}). They satisfy the following (equal $\tau$) 
Poisson bracket algebra 
\ba \nl 
\{T_0(\ul \sigma_1),T_0(\ul \sigma_2)\}&=&0,\h \{T_0(\ul 
\sigma_1),D^A_{\alpha}(\ul \sigma_2)\}=0 , \\ \nl \{T_0(\ul 
\sigma_1),T_{j}(\ul \sigma_2)\}&=& [T_0(\ul \sigma_1) + T_0(\ul 
\sigma_2)] \p_j \delta^p (\ul \sigma_1 - \ul \sigma_2) , \\ \nl 
\{T_{j}(\ul \sigma_1),T_{k}(\ul \sigma_2)\}&=& 
[\delta_{j}^{l}T_{k}(\ul \sigma_1) + \delta_{k}^{l}T_{j}(\ul 
\sigma_2)]\p_l\delta^p(\ul \sigma_1-\ul \sigma_2) , \\ \nl 
\{T_{j}(\ul \sigma_1),D^A_{\alpha}(\ul \sigma_2)\}&=& 
D^A_{\alpha}(\ul \sigma_1) \p_j \delta^p (\ul \sigma_1-\ul 
\sigma_2) , \\ \nl \{D^A_{\alpha}(\ul \sigma_1),D^B_{\beta}(\ul 
\sigma_2)\}&=& 2i\delta^{AB}\not{p}_{\alpha\beta} \delta^p (\ul 
\sigma_1-\ul \sigma_2) .  
\ea 
From the condition that the 
constraints must be maintained in time, i.e. \cite{D} 
\ba\label{Dc} 
\bigl \{T_0,H_0\bigr \}\approx 0,\h
\bigl \{T_j,H_0\bigr \}\approx 0, \h
\bigl \{D^A_\alpha,H_0\bigr \}\approx 0,
\ea
it follows that in the Hamiltonian $H_0$ one has to include the constraints
\ba
\nl
T^A_{\alpha}=\not p_{\alpha\beta} D^{A\beta}
\ea
instead of $D^A_{\alpha}$. This is because the Hamiltonian has to 
be first class quantity, but $D^A_{\alpha}$ are a mixture of first 
and second class constraints. $T^A_{\alpha}$ has the following 
non-zero Poisson brackets 
\ba \nl 
\{T_{j}(\ul \sigma_1),T^A_{\alpha}(\ul \sigma_2)\}&=& 
[T^A_{\alpha}(\ul \sigma_1)+T^A_{\alpha}(\ul \sigma_2)] 
\p_j \delta^p (\ul \sigma_1-\ul \sigma_2) , \\ 
\nl 
\{T^A_{\alpha}(\ul \sigma_1),T^B_{\beta}(\ul \sigma_2)\}&=& 
2i\delta^{AB}\not{p}_{\alpha\beta}T_0
\delta^p (\ul \sigma_1-\ul \sigma_2) .
\ea
In this form, our constraints are first class
and the Dirac consistency conditions 
(\ref{Dc}) (with $D^A_\alpha$ replaced by $T^A_\alpha$) are 
satisfied identically.  However, one now encounters a new problem. 
The constraints $T_0$, $T_j$ and $T^A_{\alpha}$ are not 
BFV-irreducible, i.e. they are functionally dependent:  
\ba \nl 
(\not p T^A)^{\alpha} - D^{A\alpha} T_0 = 0 .  
\ea 
It is known, 
that in this case after BRST-BFV quantization an infinite number of 
ghosts for ghosts appear, if one wants to preserve the manifest 
Lorentz invariance.  The way out consists in the introduction of 
auxiliary variables, so that the mixture of first and second class 
constraints $D^{A\alpha}$ can be appropriately covariantly 
decomposed into first class constraints and second class ones. To 
this end, here we will use the auxiliary harmonic variables 
introduced in \cite{Sok} and \cite{NissP}.  These are $u_{\mu}^a$ 
and $v_{\alpha}^{\pm}$, where superscripts $a=1,...,8$ and $\pm$ 
transform under the 'internal' groups $SO(8)$ and $SO(1,1)$ 
respectively. The just introduced variables are constrained by the 
following orthogonality conditions 
\ba\nl 
u_{\mu}^a u^{b\mu}=C^{ab}, \h 
u_{\mu}^{\pm}u^{a\mu}=0, \h 
u_{\mu}^{+} u^{-\mu}=-1, 
\ea 
where 
\ba\nl 
u_{\mu}^{\pm}=v_{\alpha}^\pm \sigma_{\mu}^{\alpha\beta}v_{\beta}^\pm , 
\ea 
$C^{ab}$ is the 
invariant metric tensor in the relevant representation space of 
$SO(8)$ and $(u^\pm)^2=0$ as a consequence of the Fierz identity 
for the 10-dimensional $\sigma$-matrices. We note that $u^a_\nu$ and
$v_{\alpha}^\pm$ do not depend on $\us$.

Now we have to ensure that our dynamical system does not depend on
arbitrary rotations of the auxiliary variables $(u_{\mu}^a$, $u_{\mu}^\pm)$.
It can be done by introduction of first class constraints, which
generate these transformations
\ba\nl
I^{ab}&=&-(u_{\nu}^a p^{b\nu}_u - u_{\nu}^b p^{a\nu}_u +
\frac{1}{2}v^+\sigma^{ab}p_v^+ +\frac{1}{2}v^-\sigma^{ab}p_v^-),
\h
\sigma^{ab}=u_{\mu}^a u_{\nu}^b \sigma^{\mu\nu},
\\
\label{ncon}
I^{-+}&=&-\frac{1}{2}(v_{\alpha}^+ p_v^{+\alpha} -
v_{\alpha}^- p_v^{-\alpha}),
\\ \nl
I^{\pm a}&=&-(u_{\mu}^\pm p_u^{a\mu} +
\frac{1}{2}v^\mp \sigma^\pm \sigma^a p_v^\mp) ,
\h
\sigma^\pm =u^\pm_\nu \sigma^\nu,
\h
\sigma^a = u^a_\nu \sigma^\nu .
\ea
In the above equalities, $p_u^{a\nu}$ and $p_v^{\pm \alpha}$ are the momenta
canonically conjugated to $u^a_\nu$ and $v^\pm_\alpha$.

The newly introduced constraints (\ref{ncon}) obey the following Poisson
bracket algebra
\ba\nl
\{I^{ab},I^{cd}\}&=&C^{bc}I^{ad}-C^{ac}I^{bd}+C^{ad}I^{bc}-C^{bd}I^{ac},
\\ \nl
\{I^{-+},I^{\pm a}\}&=&\pm I^{\pm a},
\\ \nl
\{I^{ab},I^{\pm c}\}&=&C^{bc}I^{\pm a}-C^{ac}I^{\pm b},
\\ \nl
\{I^{+a},I^{-b}\}&=&C^{ab}I^{-+} + I^{ab} .
\ea
This algebra is isomorphic to the $SO(1,9)$ algebra: $I^{ab}$ generate $SO(8)$
rotations, $I^{-+}$ is the generator of the subgroup $SO(1,1)$ and
$I^{\pm a}$ generate the transformations from the coset
$SO(1,9)/SO(1,1)\times SO(8)$.

Now we are ready to separate $D^{A\alpha}$ into first and second
class constraints in a Lorentz-covariant form. This separation is given
by the equalities \cite{NissPS}:
\ba\label{rev}
D^{A\alpha} &=&\frac{1}{p^+}\bigl [(\sigma^a v^+)^\alpha D^A_a +
(\not p \sigma^+ \sigma^a v^-)^\alpha G^A_a \bigr ],
\h
p^+ = p^\nu u^+_\nu,
\\ \nl
D^{Aa} &=&(v^+ \sigma^a \not p )_\beta D^{A\beta},
\h
G^{Aa} = \frac{1}{2}(v^- \sigma^a \sigma^+)_\beta D^{A\beta}.
\ea
Here $D^{Aa}$ are first class constraints and $G^{Aa}$ 
are second class ones:
\ba\nl
\{D^{Aa}(\us_1),D^{Bb}(\us_2)\}&=&
-2i\delta^{AB}C^{ab}p^+ T_0 \delta^p(\us_1 - \us_2) \\ 
\nl
\{G^{Aa}(\us_1),G^{Bb}(\us_2)\}&=&i\delta^{AB}C^{ab}p^+ \da .
\ea
It is convenient to pass from second class constraints $G^{Aa}$ 
to first class constraints $\hat G^{Aa}$, without changing the 
actual degrees of freedom \cite{NissPS}, \cite{EM} :  
\ba\nl 
G^{Aa} \rightarrow \hat G^{Aa} = G^{Aa} + (p^+)^{1/2} \Psi^{Aa} 
\h \Rightarrow \h 
\{\hat G^{Aa}(\us_1),\hat G^{Bb}(\us_2)\} = 0 ,
\ea
where $\Psi^{Aa}(\us)$ are fermionic ghosts which abelianize our second class
constraints as a consequence of the Poisson bracket relation
\ba\nl
\{\Psi^{Aa}(\us_1),\Psi^{Bb}(\us_2)\} = -i\delta^{AB}C^{ab}\da .
\ea

It turns out, that the constraint algebra is much more simple, if we work
not with $D^{Aa}$ and $\hat G^{Aa}$ but with $\hat T^{A\alpha}$ given by 
\ba\nl 
\hat T^{A\alpha} 
&=&(p^+)^{-1/2}\bigl [(\s^a v^+)^\alpha D^A_a + 
(\not p \s^+ \s^a v^-)^\alpha \hat G^A_a \bigr ] \\ 
\nl 
&=&(p^+)^{1/2}D^{A\alpha} + (\not p \s^+ \s^a v^-)^\alpha \Psi^A_a .  
\ea
After the introduction of the auxiliary fermionic variables 
$\Psi^{Aa}$, we have to modify some of the constraints, to preserve 
their first class property. Namely $T_j$, $I^{ab}$ and $I^{-a}$ 
change as follows 
\ba\nl 
\hat T_j &=& T_j + \frac{i}{2}C^{ab}\sum_A\Psi^A_a \p_j \Psi^A_b , \\ 
\nl 
\hat I^{ab} &=& I^{ab} + J^{ab}, \hspace{.3cm} 
J^{ab}=\int d^p \s j^{ab}(\us),
\hspace{.3cm}
j^{ab}=\frac{i}{4}(v^-\s_c\s^{ab}\s^+\s_d v^-)
\sum_A\Psi^{Ac}\Psi^{Ad},
\\ \nl
\hat I^{-a}&=&I^{-a} + J^{-a},
\h
J^{-a}=\int d^p\s j^{-a}(\us),
\h
j^{-a}=-(p^+)^{-1}j^{ab}p_b .
\ea
As a consequence, we can write down the Hamiltonian for the considered model
in the form:
\ba\nl
H=\int d^p\s\bigl [\la^0 T_0(\us)+\la^j \hat T_j(\us)+
\sum_A\la^{A\alpha}\hat T^A_\alpha(\us)\bigr ] +
\\ \nl
\la_{ab}\hat I^{ab}+\la_{-+}I^{-+}+\la_{+a}I^{+a}+\la_{-a}\hat I^{-a} .
\ea
The constraints entering $H$ are all first class, irreducible and Lorentz-
covariant. Their algebra reads (only the non-zero Poisson brackets are
written):
\ba\nl
\{T_0(\us_1),\hat T_j(\us_2)\}&=&\bigl (T_0(\us_1)+T_0(\us_2)\bigr )\p_j \da ,
\\ \nl
\{\hat T_j(\us_1),\hat T_k(\us_2)\}&=&
\bigl (\delta_j^l \hat T_k(\us_1)+\delta_k^l \hat T_j(\us_2)\bigr )\p_l \da ,
\\ \nl
\{\hat T_j(\us_1),\hat T^A_\alpha(\us_2)\}&=&
\bigl (\hat T^A_\alpha(\us_1)+
\frac{1}{2}\hat T^A_\alpha(\us_2)\bigr )\p_j \da ,
\\ \nl
\{\hat T^A_\alpha(\us_1),\hat T^B_\beta(\us_2)\}&=&
i\delta^{AB}\s^+_{\alpha\beta}T_0 \da ,
\\ \nl
\{I^{-+},\hat T^A_\alpha \}&=&\frac{1}{2}\hat T^A_\alpha ,
\h
\{\hat I^{-a},\hat T^A_\alpha \}=(2p^+)^{-1}\bigl [
p^a\hat T^A_\alpha + (\s^+ \s^{ab}v^-)_\alpha\Psi^A_b T_0 \bigr ] ,
\\ \nl
\{\hat I^{ab},\hat I^{cd}\}&=&C^{bc}\hat I^{ad}-C^{ac}\hat I^{bd}+
C^{ad}\hat I^{bc}-C^{bd}\hat I^{ac} ,
\\ \nl
\{I^{-+},I^{+a}\}&=&I^{+a} ,
\h
\{I^{-+},\hat I^{-a}\}=-\hat I^{-a} ,
\\ \nl
\{\hat I^{ab},I^{+c}\}&=&C^{bc}I^{+a}-C^{ac}I^{+b} ,
\h
\{\hat I^{ab},\hat I^{-c}\}=C^{bc}\hat I^{-a}-C^{ac}\hat I^{-b} ,
\\ \nl
\{I^{+a},\hat I^{-b}\}&=&C^{ab}I^{-+} + \hat I^{ab} ,
\\ \nl
\{\hat I^{-a},\hat I^{-b}\}&=&-\int d^p\s(p^+)^{-2}j^{ab}T_0 .
\ea

Having in mind the above algebra, one can construct the corresponding 
BRST charge $\Omega$ \cite{FF} ($*$=complex conjugation) 
\ba \label{O} 
\Omega = \Omega^{min}+\pi_M \bar {\cal P}^M , \h 
\{\Omega,\Omega\} = 0 , \h \Omega^* = \Omega , 
\ea 
where $M=0,j,A\alpha,ab,-+,+a,-a$.  $\Omega^{min}$ in (\ref{O}) can 
be written as 
\ba\nl 
\Omega^{min}&=&\Omega^{brane}+\Omega^{aux} , 
\\ \nl 
\Omega^{brane}&=&\int d^p\s\{T_0\eta^0+\hat T_j\eta^j+
\sum_A\hat T^A_\alpha \eta^{A\alpha} + 
{\cal P}_0 [(\p_j\eta^j)\eta^0 + (\p_j\eta^0)\eta^j ] + 
\\ \nl &+&{\cal P}_k(\p_j\eta^k)\eta^j + 
\sum_A{\cal P}^A_\alpha [\eta^j\p_j\eta^{A\alpha} -
\frac{1}{2}\eta^{A\alpha}\p_j\eta^j ] -
\frac{i}{2}{\cal P}_0
\sum_A\eta^{A\alpha}\s^+_{\alpha\beta}\eta^{A\beta}\} ,
\\ \nl
\Omega^{aux}&=&
\hat I^{ab}\eta_{ab}+I^{-+}\eta_{-+}+I^{+a}\eta_{+a}+\hat I^{-a}\eta_{-a} 
\\ \nl
&+&({\cal P}^{ac}\eta^{b.}_{.c}-{\cal P}^{bc}\eta^{a.}_{.c} +
2{\cal P}^{+a}\eta^b_+ + 2{\cal P}^{-a}\eta^b_-)\eta_{ab} 
\\ \nl
&+&({\cal P}^{+a}\eta_{+a}-{\cal P}^{-a}\eta_{-a})\eta_{-+} +
({\cal P}^{-+}\eta^a_- + {\cal P}^{ab}\eta_{-b})\eta_{+a} 
\\ \nl
&+&\frac{1}{2}\int d^p\s\{
\sum_A{\cal P}^A_\alpha\eta^{A\alpha}\eta_{-+} +
(p^+)^{-1}\sum_A[p^a{\cal P}^A_\alpha -
(\s^+\s^{ab}v^-)_\alpha\Psi^A_b {\cal P}_0]
\eta^{A\alpha}\eta_{-a}
\\ \nl
&-&(p^+)^{-2}j^{ab}{\cal P}_0\eta_{-b}\eta_{-a}\} .
\ea
These expressions for $\Omega^{brane}$ and $\Omega^{aux}$ show that we have
found a set of constraints which ensure the first rank property of the model.

$\Omega^{min}$ can be represented also in the form
\ba\nl
\Omega^{min}=\int d^p\s\bigl [\bigl (T_0+\frac{1}{2}T_0^{gh}\bigr )\eta^0
+\bigl (\hat T_j+\frac{1}{2}T_j^{gh}\bigr )\eta^j+
\sum_A
\bigl (\hat T^A_\alpha +\frac{1}{2}T_\alpha^{A gh}\bigr )
\eta^{A\alpha}\bigr ]
\\ \nl
+\bigl (\hat I^{ab}+\frac{1}{2}I^{ab}_{gh}\bigr )\eta_{ab}
+\bigl (I^{-+}+\frac{1}{2}I^{-+}_{gh}\bigr )\eta_{-+}
+\bigl (I^{+a}+\frac{1}{2}I^{+a}_{gh}\bigr )\eta_{+a}
+\bigl (\hat I^{-a}+\frac{1}{2}I^{-a}_{gh}\bigr )\eta_{-a}
\\ \nl
+\int d^p\s\p_j\Bigl (\frac{1}{2}{\cal P}_k\eta^k\eta^j
+\frac{1}{4}
\sum_A{\cal P}^A_\alpha\eta^{A\alpha}\eta^j\Bigr ) .
\ea
Here a super(sub)script $gh$ is used for the ghost part of the total gauge
generators
\ba\nl
\nl
G^{tot}=\{\Omega,{\cal P}\}=\{\Omega^{min},{\cal P}\}=G+G^{gh} .
\ea
We recall that the Poisson bracket algebras of $G^{tot}$ and $G$ coincide
for first rank systems. The manifest expressions for $G^{gh}$ are:
\ba\nl
T_0^{gh}&=&2{\cal P}_0\p_j\eta^j+\bigl (\p_j{\cal P}\bigr )\eta^j ,
\\ \nl
T_j^{gh}&=&2{\cal P}_0\p_j\eta^0+\bigl (\p_j{\cal P}_0\bigr )\eta^0+
{\cal P}_j\p_k\eta^k+{\cal P}_k\p_j\eta^k+\bigl (\p_k{\cal P}_j\bigr )\eta^k
\\ \nl
&+&\frac{3}{2}\sum_A{\cal P}^A_\alpha\p_j\eta^{A\alpha}
+\frac{1}{2}\sum_A\bigl (\p_j{\cal P}^A_\alpha\bigr )\eta^{A\alpha},
\\ \nl
T_\alpha^{A gh}&=&-\frac{3}{2}{\cal P}^A_\alpha\p_j\eta^j
-\bigl (\p_j{\cal P}^A_\alpha\bigr )\eta^j
-i{\cal P}_0\s^+_{\alpha\beta}\eta^{A\beta} +
\\ \nl
&+&\frac{1}{2}{\cal P}^A_\alpha\eta_{-+}
+(2p^+)^{-1}\Bigl [p^a{\cal P}^A_\alpha
-(\s^+\s^{ab}v^-)_\alpha\Psi^A_b{\cal P}_0\Bigr ]\eta_{-a} ,
\\ \nl
I^{ab}_{gh}&=&2\bigl ({\cal P}^{ac}\eta^{b.}_{.c}-{\cal P}^{bc}\eta^{a.}_{.c}
\bigr )+\bigl ({\cal P}^{+a}\eta^b_+-{\cal P}^{+b}\eta^a_+\bigr )+
\bigl ({\cal P}^{-a}\eta^b_--{\cal P}^{-b}\eta^a_-\bigr ) ,
\\ \nl
I^{-+}_{gh}&=&{\cal P}^{+a}\eta_{+a}-{\cal P}^{-a}\eta_{-a}+
\frac{1}{2}\int d^p\s\sum_A{\cal P}^A_\alpha\eta^{A\alpha} ,
\\ \nl
I^{+a}_{gh}&=&2{\cal P}^{+b}\eta^{a.}_{.b}-{\cal P}^{+a}\eta_{-+}+
{\cal P}^{-+}\eta^a_-+{\cal P}^{ab}\eta_{-b} ,
\\ \nl
I^{-a}_{gh}&=&2{\cal P}^{-b}\eta^{a.}_{.b}+{\cal P}^{-a}\eta_{-+}-
{\cal P}^{-+}\eta^a_++{\cal P}^{ab}\eta_{+b}+
\\ \nl
&+&\int d^p\s\Bigl \{(2p^+)^{-1}
\sum_A\Bigl [p^a{\cal P}^A_\alpha-
(\s^+\s^{ab}v^-)_\alpha\Psi^A_b{\cal P}_0\Bigr ]\eta^{A\alpha}-
(p^+)^{-2}j^{ab}{\cal P}_0 \eta_{-b}\Bigr \} .
\ea
Up to now, we introduced canonically conjugated ghosts
$\bigl (\eta^M,{\cal P}_M\bigr )$, $\bigl (\bar \eta_M,\bar {\cal P}^M\bigr)$
and momenta $\pi_M$ for the Lagrange multipliers $\lambda^M$ in the
Hamiltonian. They have Poisson brackets and Grassmann parity as follows
($\epsilon_M$ is the Grassmann parity of the corresponding 
constraint):  
\ba\nl 
\bigl \{\eta^M,{\cal P}_N\bigr \}&=&\delta^M_N , 
\h 
\epsilon (\eta^M)=\epsilon ({\cal P}_M)=\epsilon_M + 1 , \\ 
\nl 
\bigl \{\bar \eta_M,\bar {\cal P}^N \bigr \}&=&-(-1)^{\epsilon_M\epsilon_N} 
\delta^N_M , 
\h 
\epsilon (\bar\eta_M)=\epsilon (\bar {\cal P}^M)=\epsilon_M + 1 , 
\\ \nl 
\bigl \{\lambda^M,\pi_N\bigr \}&=&\delta^M_N ,
\h
\epsilon (\lambda^M)=\epsilon (\pi_M)=\epsilon_M .
\ea

The BRST-invariant Hamiltonian is
\ba\label{H}
H_{\tilde \chi}=H^{min}+\bigl \{\tilde \chi,\Omega\bigr \}=
\bigl \{\tilde \chi,\Omega\bigr \} ,
\ea
because from $H_{canonical}=0$ it follows $H^{min}=0$. In this
formula $\tilde \chi$ stands for the gauge fixing fermion
$(\tilde \chi^* = -\tilde \chi)$. We use the following representation
for the latter
\ba\nl
\tilde \chi=\chi^{min}+\bar\eta_M(\chi^M+\frac{1}{2}\rho_{(M)}\pi^M) ,
\h
\chi^{min}=\la^M{\cal P}_M ,
\ea
where $\rho_{(M)}$ are scalar parameters and we have separated the
$\pi^M$-dependence from $\chi^M$. If we adopt that $\chi^M$ does 
not depend on the ghosts $(\eta^M,{\cal P}_M)$ and 
$(\bar\eta_M,\bar {\cal P}^M)$, the Hamiltonian $H_{\tilde\chi}$ 
from (\ref{H}) takes the form 
\ba
\label{r} 
H_{\tilde\chi}&=&H_{\chi}^{min}+{\cal P}_M \bar {\cal P}^M - 
\pi_M(\chi^M+\frac{1}{2}\rho_{(M)}\pi^M)+ \\ \nl &+&\bar\eta_M\Bigl 
[\bigl \{\chi^M,G_N\bigr \}\eta^N +\frac{1}{2}(-1)^{\epsilon_N} 
{\cal P}_Q\bigl \{\chi^M,U^Q_{NP}\bigr \}\eta^P\eta^N \Bigr ],
\ea
where
\ba\nl
H_{\chi}^{min}=\bigl \{\chi^{min},\Omega^{min}\bigr \} ,
\ea
and generally $\bigl \{\chi^M,U^Q_{NP}\bigr \}\not=0$ as far as the structure
coefficients of the constraint algebra $U^M_{NP}$ depend on the phase-space
variables.

One can use the representation (\ref{r}) for $H_{\tilde\chi}$ to obtain the
corresponding BRST invariant Lagrangian
\ba\nl
L_{\tilde\chi}=L+L_{GH}+L_{GF} .
\ea
Here $L_{GH}$ stands for the ghost part and $L_{GF}$ - for the gauge fixing
part of the Lagrangian. If one does not intend to pass to the Lagrangian
formalism, one may restrict oneself to
the minimal sector $\bigl (\Omega^{min},\chi^{min},H_\chi^{min}\bigr )$.
In particular, this means that Lagrange multipliers are not considered as
dynamical variables anymore. 
With this particular gauge choice, $H_\chi^{min}$
is a linear combination of the total constraints
\ba\nl
H_\chi^{min}&=&H_{brane}^{min}+H_{aux}^{min}=
\\ \nl
&=&\int d^p\s\Bigl [\La^0 T_0^{tot}(\us)+\La^j T_j^{tot}(\us)+
\sum_A\La^{A\alpha} T_\alpha^{A tot}(\us)\Bigr ] +
\\ \nl
&+&\La_{ab}I^{ab}_{tot}+\La_{-+}I^{-+}_{tot}+\La_{+a}I^{+a}_{tot}+
\La_{-a}I^{-a}_{tot} ,
\ea
and we can treat here the Lagrange multipliers $\La^0,...,\La_{-a}$
as constants. Of course, this does not fix the gauge completely.
%%%%%%%%%%%%%%%%%%%%%%%%%%%%%%%%%%%%%%%%%%%%%%%%%%%%%%%%%%%%%%%%%%%%%%%%%%%%%
\section{\bf Comments and conclusions}
%%%%%%%%%%%%%%%%%%%%%%%%%%%%%%%%%%%%%%%%%%%%%%%%%%%%%%%%%%%%%%%%%%%%%%%%%%%%%
\hspace{1cm}
To ensure that the harmonics and their conjugate momenta are pure
gauge degrees of freedom, we have to consider as physical
observables only such functions on the phase space which do not
carry any $SO(1,1)\times SO(8)$ indices.  More precisely, these
functions are defined by the following expansion
\ba\nl
F(y,u,v;p_y,p_u,p_v)=\sum_{}\bigl [u_{\nu_1}^{a_1}...u_{\nu_k}^{a_k}
p_{u\nu_{k+1}}^{a_{k+1}}...p_{u\nu_{k+l}}^{a_{k+l}}\bigr ]_{SO(8)
singlet}\\ \nl
v_{\alpha_1}^+...v_{\alpha_m}^+v_{\alpha_{m+1}}^-...v_{\alpha_{m+n}}^-
p_v^{+\beta_1}...p_v^{+\beta_r}p_v^{-\beta_{r+1}}...p_v^{-\beta_{m-n+r}}\\
\nl
F_{\beta_1...\beta_{m-n+r}}^{\alpha_1...\alpha_{m+n},\nu_1...\nu_{k+l}}
(y,p_y) ,
\ea
where $(y,p_y)$ are the non-harmonic phase space conjugated pairs.

In this letter we consider a model for tensionless super 
$p$-branes with N global chiral supersymmetries in 10-dimensional 
Minkowski space-time. We show that the action is reparametrization 
and kappa-invariant. After establishing the symmetries of the action, 
we give the general solution of the classical equations of motion 
in a particular gauge. In the case of null superstrings ($p$=1) 
we find the general solution in an arbitrary gauge. Starting with a 
Hamiltonian which is a linear combination of first and mixed (first 
and second) class constraints, we succeed to obtain a new one, 
which is a linear combination of first class, BFV-irreducible and 
Lorentz-covariant constraints only. This is done with the help of 
the introduced auxiliary harmonic variables. Then we give manifest 
expressions for the classical BRST charge, the corresponding total 
constraints and BRST-invariant Hamiltonian. It turns out, that in 
the given formulation our model is a first rank dynamical system.

%%%\vspace*{1cm}
%%%{\bf Acknowledgments}
%%%
%%%The author would like to thank ........... for careful reading 
%%%of the manuscript.

\vspace*{1cm}
%%%%%%%%%%%%%%%%%%%%%%%%%%%%%%%%%%%%%%%%%%%%%%%%%%%%%%%%%%%%%%%%%%%%%%%%%%%%
{\Large{\bf Appendix}}
%%%%%%%%%%%%%%%%%%%%%%%%%%%%%%%%%%%%%%%%%%%%%%%%%%%%%%%%%%%%%%%%%%%%%%%%%%%%
\vspace*{.5cm}
\hspace{1cm}

We briefly describe here our 10-dimensional conventions. Dirac $\gamma$-
matrices obey
\ba\nl
\Gamma_\mu\Gamma_\nu+\Gamma_\nu\Gamma_\mu=2\eta_{\mu\nu}
\ea
and are taken in the representation
\ba\nl
\Gamma^\mu = \left(\begin{array}{cc}0&(\s^\mu)_\alpha^{\dot\beta}\\
(\tilde\s^\mu)^\beta_{\dot\alpha}&0\end{array}\right)
\h .
\ea
$\Gamma^{11}$ and charge conjugation matrix $C_{10}$ are given by
\ba\nl
\Gamma^{11}=\Gamma^0\Gamma^1 ... \Gamma^9 =
\left(\begin{array}{cc}\delta_\alpha^\beta &0\\
0&-\delta_{\dot\alpha}^{\dot\beta}\end{array}\right)
\h ,
\ea
\ba \nl
C_{10}=\left(\begin{array}{cc}0&C^{\alpha\dot\beta}\\
(-C)^{\dot\alpha\beta}&0\end{array}\right)
\h ,
\ea
and the indices of right and left Majorana-Weyl fermions are raised as
\ba\nl
\psi^\alpha=C^{\alpha\dot\beta}\psi_{\dot\beta}
\h ,\h
\phi^{\dot\alpha}=(-C)^{\dot\alpha\beta}\phi_\beta .
\ea

We use $D=10$ $\s$-matrices with undotted indices
\ba\nl
(\s^\mu)^{\alpha\beta}=C^{\alpha\dot\alpha}(\tilde\s^\mu)_{\dot\alpha}^\beta
\h ,\h
(\s^\mu)_{\alpha\beta}=(-C)^{-1}_{\beta\dot\beta}(\s^\mu)_\alpha^{\dot\beta}
\h ,
\ea
and the notation
\ba\nl
\s^{\mu_1...\mu_n}\equiv\s^{[\mu_1}...\s^{\mu_n]}
\ea
for their antisymmetrized products.

From the corresponding properties of $D=10$ $\gamma$-matrices, it follows:
\ba\nl
(\s^\mu)_{\alpha\gamma}(\s^\nu)^{\gamma\beta}+
(\s^\nu)_{\alpha\gamma}(\s^\mu)^{\gamma\beta}=
-2\delta_\alpha^\beta\eta^{\mu\nu}\h ,
\\ \nl
(\s_{\mu_1 ... \mu_{2s+1}})^{\alpha\beta}=
(-1)^s(\s_{\mu_1 ... \mu_{2s+1}})^{\beta\alpha}\h ,
\\ \nl
\s^\mu\s^{\nu_1 ... \nu_n}=\s^{\mu\nu_1 ... \nu_n}+
\sum_{k=1}^{n}(-1)^k\eta^{\mu\nu_k}
\s^{\nu_1 ... \nu_{k-1}\nu_{k+1} ... \nu_n}.
\ea

The Fierz identity for the $\s$-matrices reads:
\ba\nl
(\s_\mu)^{\alpha\beta}(\s^\mu)^{\gamma\delta}+
(\s_\mu)^{\beta\gamma}(\s^\mu)^{\alpha\delta}+
(\s_\mu)^{\gamma\alpha}(\s^\mu)^{\beta\delta} = 0 .
\ea

\vspace*{.5cm}

%%%%%%%%%%%%%%%%%%%%%%%%%%%%%%%%%%%%%%%%%%%%%%%%%%%%%%%%%%%%%%%%%%%%%%%%%%%%

\end{document}